\PassOptionsToPackage{svgnames}{xcolor}
\documentclass[a4paper,twocolumn,11pt,unpublished]{quantumarticle}
\pdfoutput=1

\usepackage[utf8]{inputenc}
\usepackage[T1]{fontenc}
\usepackage{amsmath}
\usepackage{amssymb}
\usepackage{amsfonts}
\usepackage{hyperref}
\usepackage{booktabs}
\usepackage{nicefrac}
\usepackage{microtype}
\usepackage{algpseudocode}
\usepackage{listings}
\usepackage{graphicx}
\usepackage{subcaption}
\usepackage{enumitem}
\usepackage{braket}
\usepackage{tikz}
\usetikzlibrary{fit}
\usepackage[numbers]{natbib}
\usepackage{orcidlink}

\newcommand{\ad}{a^\dagger} 

\lstdefinelanguage{openqasm}{
  keywords=[1]{
    OPENQASM, include, defcalgrammar, cal, defcal, gate, def, extern,
    box, let, break, continue, if, else, end, return, for, while, in,
    pragma, input, output, const, readonly, mutable,
    qreg, creg, qubit, qumode, bit, int, uint, float, angle, bool, duration,
    stretch, complex, array, void, size, sizeof
  },
  keywords=[2]{
    measure, reset, barrier, delay, durationof, ctrl, negctrl, 
    inv, pow, measure_x, measure_n
  },
  keywords=[3]{
    bit, qubit, int, uint, float, angle, bool, duration, stretch, complex, array, void
  },
  keywords=[4]{
    true, false, pi, tau, euler, im
  },
  sensitive=true,
  comment=[l]{//},
  morecomment=[s]{/*}{*/},
  morestring=[b]",
  morestring=[b]',
  basicstyle=\scriptsize\ttfamily,
  breaklines=true,
  keepspaces=true,
  showspaces=false,
  showstringspaces=false,
  showtabs=false,
  tabsize=2
}

\lstdefinelanguage{jaqal}{
  keywords=[1]{
    from, usepulses, subcircuit, register, map, let, macro, loop
  },
  sensitive=true,
  comment=[l]{//},
  morecomment=[s]{/*}{*/},
  morestring=[b]",
  morestring=[b]',
  basicstyle=\scriptsize\ttfamily,
  breaklines=true,
  keepspaces=true,
  showspaces=false,
  showstringspaces=false,
  showtabs=false,
  tabsize=2
}

\begin{document}

\title{Hybridlane: A Software Development Kit for Hybrid Continuous-Discrete Variable Quantum Computing}

\author{Jim Furches \orcidlink{0009-0008-2852-4208}}
\email{james.furches@pnnl.gov}
\affiliation{Pacific Northwest National Laboratory, Richland, WA 99354, USA}

\author{Timothy J. Stavenger \orcidlink{0000-0002-4270-5952}}
\email{timothy.stavenger@pnnl.gov}
\affiliation{Pacific Northwest National Laboratory, Richland, WA 99354, USA}

\author{Carlos Ortiz Marrero \orcidlink{0000-0001-8302-1322}}
\email{carlos.ortiz.marrero@colostate.edu}
\affiliation{Pacific Northwest National Laboratory, Richland, WA 99354, USA}
\affiliation{Colorado State University, Fort Collins, CO 80523, USA}

\maketitle

\begin{abstract}
Hybrid quantum computing systems that combine discrete-variable qubits with continuous-variable qumodes offer promising advantages for quantum simulation, error correction, and sensing applications. However, existing quantum software frameworks lack native support for expressing and manipulating hybrid circuits, forcing developers to work with fragmented toolchains or rely on simulation-coupled representations that limit scalability. We present Hybridlane, an open-source software development kit providing a unified frontend for hybrid continuous-discrete variable quantum computing. Hybridlane introduces automatic wire type inference to distinguish qubits from qumodes without manual annotations, enabling compile-time validation of circuit correctness. By decoupling gate semantics from matrix representations, Hybridlane can describe wide and deep circuits with minimal memory consumption and without requiring simulation. The framework implements a comprehensive library of hybrid gates and decompositions following established instruction set architectures, while remaining compatible with PennyLane's extensive qubit algorithm library. Furthermore, it supports multiple backends including classical simulation with Bosonic Qiskit and hardware compilation to Sandia National Laboratories' QSCOUT ion trap. We demonstrate Hybridlane's capabilities through bosonic quantum phase estimation and ion trap calibration workflows.
\end{abstract}

\section{Introduction}
\label{sec:introduction}

Quantum computing platforms can be broadly classified into two paradigms: discrete-variable (DV) systems operating on qubits (or qudits) with a finite number of basis states and continuous-variable (CV) systems operating on qumodes with infinite-dimensional Hilbert spaces \cite{liu2026hybrid}. Each paradigm offers distinct advantages: qubits provide robust digital encoding and established error correction protocols, while qumodes offer large Hilbert spaces enabling compact state representations, bosonic error correction codes, and analog quantum simulation~\cite{gottesman2001encoding,lloyd1999quantum}.

Recent theoretical and experimental advances demonstrate that \emph{hybrid} platforms combining qubits and qumodes can leverage the complementary strengths of both paradigms. Crane et al.~\cite{crane2024hybrid} showed that hybrid oscillator-qubit processors enable efficient simulation of fermions, bosons, and gauge fields with polynomial resource scaling. Liu et al.~\cite{liu2026hybrid} formalized hybrid instruction set architectures and abstract machine models, establishing a theoretical foundation for programming these systems. Kemper et al.~\cite{kemper2025hybrid} provide a comprehensive guide to the potential utility of hybrid CV-DV quantum computing, outlining applications spanning physics, chemistry, and computer science domains. Furthermore, multiple near-term hardware platforms naturally realize hybrid systems, including trapped-ion processors where ionic qubits couple to collective motional modes~\cite{clark2021qscout,araz2024hybrid}, superconducting circuits where transmon qubits interact with cavity oscillators~\cite{blais2021circuit}, and photonic systems combining path-encoded qubits with photon-number or coherent-state qumodes.

The quantum software ecosystem has been essential to investigating the utility of qubit-based quantum computers by permitting simulation, resource estimation, and hardware execution. Currently, the ecosystem for CV-DV quantum is comparatively less mature and fragmented. The first issue is that many quantum libraries are not designed for hybrid computing; some lack support for qumodes \cite{qiskit,cirq,sivarajah2021pytket,killoran2019strawberry,qualtran} or do not permit heterogeneous quantum circuits \cite{killoran2019strawberry,bergholm2018pennylane}. Secondly, of the libraries that do support the hybrid CV-DV paradigm, each is missing key features for general-purpose quantum circuits, such as gate-level programming \cite{qutip5,chen2025genesis} and symbolic manipulation \cite{stavenger2022bosonic}.

Therefore, to the best of our knowledge, no existing software development kit provides all the features one might desire for a CV-DV quantum SDK: heterogenous quantum registers, support for expressing general-purpose, hybrid quantum programs in a scalable and backend-independent manner, and execution on multiple backends (simulator or hardware). This ecosystem gap impedes algorithm development and prevents the kind of rapid prototyping with unified workflows that accelerated progress in qubit-based quantum computing. Recent benchmarking efforts such as HyQBench~\cite{mohapatra2025hyqbench} further highlight the need for mature tooling by demonstrating the resource advantages of hybrid circuits across diverse applications.

To address this ecosystem gap, we introduce Hybridlane, a software development kit providing a unified frontend for hybrid continuous-discrete variable quantum computing. Hybridlane has several key features that we believe are useful to hybrid CV-DV quantum programming:

\begin{itemize}
\item \textbf{Native qumodes and wire type inference}: Circuit wires in Hybridlane are statically typed as $\{\mathtt{qubit}, \mathtt{qumode}, \mathtt{qudit}(d)\}$ to allow compile-time circuit validation. But, these types are determined automatically prior to execution through a type inference algorithm (Section \ref{sec:wire_types}), preventing manual type annotations that would inconvenience the user.

\item \textbf{Hybrid gate library and decompositions}: Hybridlane implements most of the gates in Liu et al.~\cite{liu2026hybrid}, spanning the phase-space, Fock-space, and sideband instruction set architectures (ISAs). In addition, many gate decomposition identities from Liu et al.~\cite{liu2026hybrid} and Crane et al.~\cite{crane2024hybrid} are incorporated into PennyLane's experimental graph decomposition system, enabling algorithm development, quantum resource estimation, and programming across diverse hybrid hardware platforms. The gates and decomposition identities are enumerated in Section \ref{sec:gates}.

\item \textbf{CV measurement framework}: Hybridlane extends PennyLane's measurement model to support infinite-spectrum CV observables (e.g., $\hat{n}$, $\hat{x}$) through spectrum functions rather than finite eigenvalue arrays, and tracks measurement bases (Fock, homodyne) per wire (Section~\ref{sec:measurements}).

\item \textbf{Multi-backend support}: Leveraging PennyLane's \texttt{Device} interface, Hybridlane provides a unified abstraction for backend execution. Current implementations include classical simulation via Bosonic Qiskit~\cite{stavenger2022bosonic} (Section~\ref{subsec:bosonic-qiskit}) and hardware compilation to Sandia National Laboratories' QSCOUT ion trap~\cite{clark2021qscout} (Section~\ref{subsec:sandia}), with the architecture enabling future backend integrations.

\item \textbf{OpenQASM extensions}: Proposed extensions to OpenQASM 3.0 introduce \texttt{qumode}, \texttt{measure\_x}, and \texttt{measure\_n} keywords with a standard gate library (\texttt{cvstdgates.inc}), enabling hybrid circuit serialization and facilitating future backend integrations (Section~\ref{subsec:openqasm}).

\item \textbf{PennyLane compatibility}: Because Hybridlane extends PennyLane, hybrid quantum programs can leverage PennyLane's comprehensive library of qubit gates and algorithms, which are used in many CV-DV routines \cite{liu2026hybrid}. Additionally, the design of PennyLane enables the creation of reusable hybrid circuits and algorithms inside Hybridlane, which is hard to achieve with existing software.

\end{itemize}

\begin{figure*}[t]
    \centering
    \includegraphics[width=\linewidth]{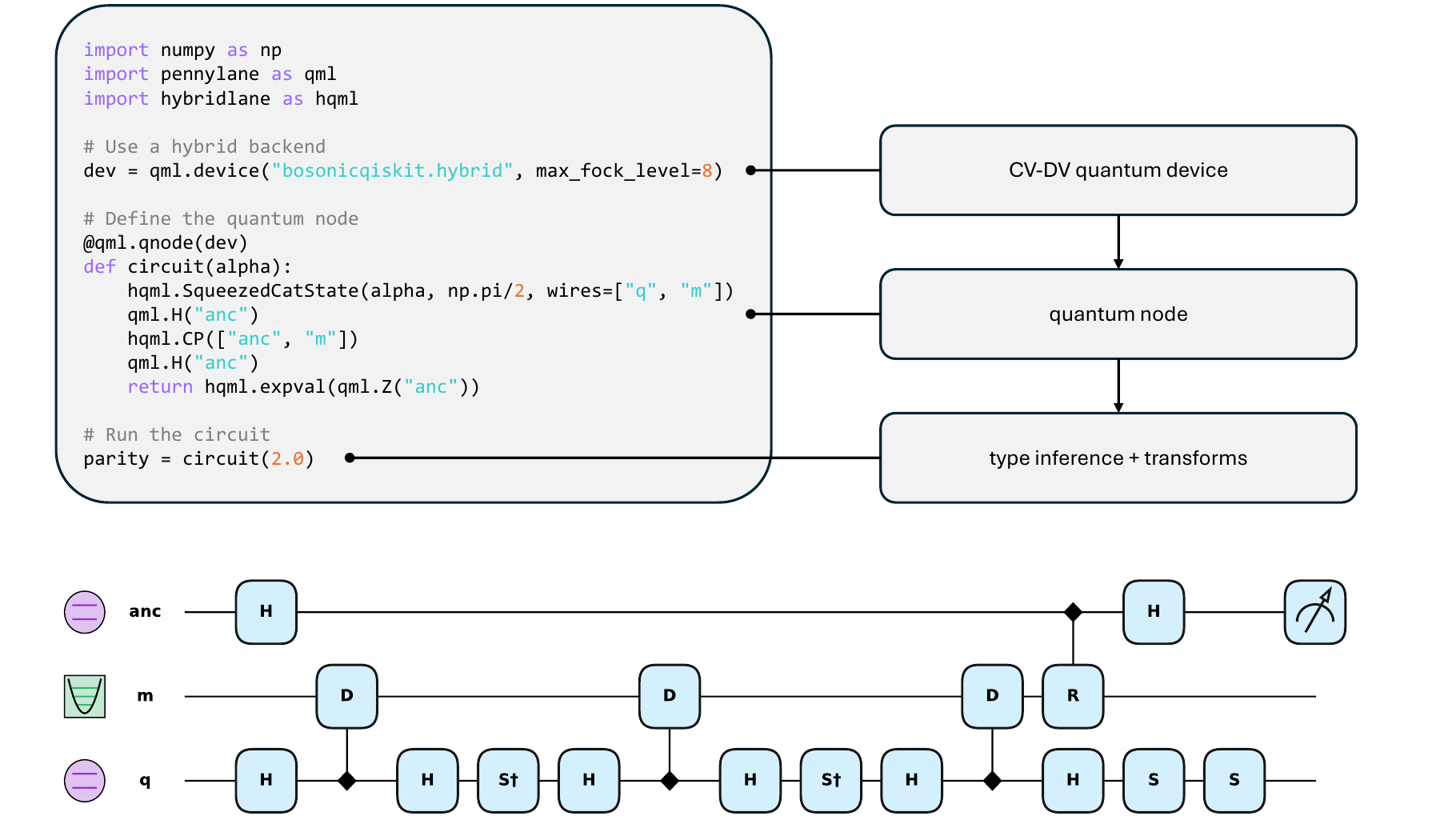}
    \caption{Basic example of defining a hybrid CV-DV quantum program in Hybridlane, following the standard PennyLane syntax. Type inference and circuit transforms allow targeting the quantum node to specific CV-DV devices.}
    \label{fig:hybridlane_overview}
\end{figure*}

Hybridlane is open source (BSD-2-Clause), actively developed on GitHub\footnote{\url{https://github.com/pnnl/hybridlane}}, with API documentation and demos on the GitHub Pages website\footnote{\url{https://pnnl.github.io/hybridlane}}.

The remainder of this paper is organized as follows. Section~\ref{sec:background} provides background on hybrid CV-DV quantum computing and positions Hybridlane within the quantum software ecosystem. Section~\ref{sec:design} presents Hybridlane's design and architecture, including the wire type system and automatic type inference algorithm and the gate library and current decomposition framework. Section~\ref{sec:backends} details backend integration, including Bosonic Qiskit simulation and export to JAQAL and OpenQASM intermediate representations. Section~\ref{sec:applications} demonstrates Hybridlane through two applications: quantum phase estimation, and ion trap calibration.

\section{Background}
\label{sec:background}

In this section, we provide essential background on hybrid continuous-discrete variable quantum computing and survey existing software frameworks, highlighting gaps that Hybridlane aims to address. We begin by briefly reviewing the aspects of CV-DV quantum computing relevant to software frameworks, but we refer the reader to Liu et al. \cite{liu2026hybrid} for a comprehensive introduction and survey of the literature on hybrid CV-DV computing.

\subsection{CV-DV Quantum Computation}

Qubits are two-level quantum systems ($\mathcal{H} = \mathbb{C}^2$) with discrete measurement outcomes $\{0,1\}$ corresponding to the computational (Z) basis $\{\ket{0}, \ket{1}\}$. Measurements in other bases can be performed by first applying a basis rotation prior to measurement in the computational basis; thus, in a quantum software framework, it suffices to provide only Z-basis measurements. Any qubit algorithm can be performed through a sequence of unitary gates drawn from a universal gate set, augmented with ancilla qubits, mid-circuit measurements, and classical feedforward. Notably, qubit quantum computers have a restricted, non-universal Clifford gate set (Hadamard, Phase, CNOT) that can be efficiently simulated on a classical computer; adding a single non-Clifford gate (such as $T$) promotes this to a universal gate set.

While qubits have a natural digital structure, qumodes describe the quantum states of physical oscillators, such as electromagnetic field modes in a cavity or motional modes of trapped ions, whose continuous nature demands a fundamentally different mathematical framework. Qumodes are quantum harmonic oscillators with infinite-dimensional Hilbert spaces. Unlike qubits, qumodes require two inequivalent computational bases: the Fock basis $\mathcal{H} = \text{span}\{\ket{n}\}_{n \in \mathbb{N}_0}$ with discrete photon number states, and the position basis $\mathcal{H} = \text{span}\{\ket{x}\}_{x \in \mathbb{R}}$ representing continuous field quadratures. While these bases are related by a unitary transformation involving Hermite-Gaussian functions, both are essential in a software framework because they correspond to distinct physical measurement primitives. The notable Gaussian qumode gates (Table~\ref{tab:cv-gates}) include Displacement $D(\alpha)$, Squeezing $S(\zeta)$, Phase Rotation $R(\theta)$, and Beamsplitter $BS(\theta, \phi)$; these preserve Gaussian states and can be efficiently simulated classically. Non-Gaussian operations such as the Kerr gate $K(\kappa)$ or photon number resolving measurements are required for universal CV computation. Qumodes support multiple measurement types: Fock measurement (photon counting: $\hat{n} \to \mathbb{N}_0$), Homodyne detection (quadrature measurement: $\hat{x}_\phi \to \mathbb{R}$), and Heterodyne detection (simultaneous noisy measurement of both quadratures: $\to \mathbb{C}$).

Hybrid gates (Table~\ref{tab:hybrid-gates}) couple qubits and qumodes, enabling energy exchange and conditional operations~\cite{liu2026hybrid}. For example, the Jaynes-Cummings gate $JC(\theta, \phi)$ exchanges qubit-qumode excitations, modeling atom-cavity coupling, and the Conditional Rotation gate $CR(\theta)$ implements dispersive coupling, imparting a qubit phase proportional to photon number. As hybrid gates are native to many platforms and are neither Clifford nor Gaussian, they enable applications like bosonic codes~\cite{gottesman2001encoding}, quantum sensing~\cite{ivanov2016highprecision}, and bosonic simulation~\cite{crane2024hybrid}.

\subsection{Existing Software Ecosystem}

\begin{table*}[th]
\caption{Evaluation of quantum software frameworks against identified architectural requirements. Hybridlane uniquely fulfills all four criteria for native, backend-independent CV-DV computing.}
\label{tab:ecosystem-comparison}
\centering
\small
\begin{tabular}{lccccc}
\toprule
\textbf{Requirement} & \textbf{Qiskit} & \textbf{PennyLane} & \textbf{Strawberry} & \textbf{Bosonic} & \textbf{Hybridlane} \\
& \cite{qiskit} & \cite{bergholm2018pennylane} & \textbf{Fields} \cite{killoran2019strawberry} & \textbf{Qiskit} \cite{stavenger2022bosonic} & (Ours) \\
\midrule
1. Native CV-DV Types & DV Only & \checkmark\textsuperscript{a} & CV Only & \checkmark\textsuperscript{b} & \checkmark \\
2. CV-DV Measurements & \texttimes & \checkmark\textsuperscript{c} & \checkmark & \texttimes\textsuperscript{d} & \checkmark \\
3. Reusable Subroutines & \checkmark & \checkmark & \checkmark & \texttimes & \checkmark \\
4. Backend Independence & \checkmark & \checkmark & \checkmark & \texttimes\textsuperscript{e} & \checkmark \\
\bottomrule
\end{tabular}

\vspace{0.5em}
\footnotesize
\begin{flushleft}
\textsuperscript{a} PennyLane currently lacks devices supporting simultaneous DV and CV operations. \\
\textsuperscript{b} Represents qumodes internally as qubits in Qiskit’s IR; lacks a distinct qumode abstraction. \\
\textsuperscript{c} Core measurement API assumes finite eigenvalue spectra; CV measurements require plugin-specific handling that bypasses the standard framework. \\
\textsuperscript{d} Bosonic Qiskit requires manual construction of Fock-space truncated observables. \\
\textsuperscript{e} Tight coupling between representation and Fock-space simulation prevents hardware-agnostic dispatch.
\end{flushleft}
\end{table*}

Drawing from the hybrid instruction set architecture formalized by Liu et al.~\cite{liu2026hybrid} and the requirements of multi-backend quantum programming, we identify four essential architectural requirements for a hybrid CV-DV frontend:

\begin{enumerate}
    \item \textbf{Native CV-DV Types}: The ability to address both DV and CV degrees of freedom within a single quantum circuit.

    \item \textbf{CV-DV Measurements}: Support for CV observables, including those with unbounded spectra (e.g., $\hat{x}$, $\hat{n}$) and measurement in different computational bases (e.g., photon number readout, homodyne measurement).

    \item \textbf{Reusable Subroutines}: The ability to define modular templates that encapsulate both DV and CV gates for scalable circuit construction.

    \item \textbf{Backend Independence}: A symbolic representation that allows the same circuit to be dispatched to simulators or hardware.
\end{enumerate}

We evaluate current SDKs against these criteria (see Table \ref{tab:ecosystem-comparison}). Qubit-centric frameworks such as \textbf{Qiskit} \cite{qiskit} and \textbf{pytket} \cite{sivarajah2021pytket} utilize fixed-dimension qubit registers, which inherently exclude CV states. While \textbf{Cirq} \cite{cirq} provides a \texttt{Qid} type for general quantum objects, it enforces a finite integer dimension, making it incompatible with the infinite-dimensional Hilbert space of a qumode. Similarly, \textbf{Qualtran} \cite{qualtran} allows finite-dimensional custom data types (\texttt{QDtype}) as abstractions for decomposition into qubits.

Among bosonic-capable libraries, \textbf{Strawberry Fields} \cite{killoran2019strawberry}, \textbf{MrMustard} \cite{mrmustard2024}, and \textbf{Piquasso} \cite{piquasso2025} are specialized for CV optics but lack the primitives to represent qubit registers or hybrid interactions. \textbf{QuTiP} \cite{qutip5} offers robust simulation for hybrid systems via tensor products of Fock spaces and spin systems, yet it does not adhere to the gate-based quantum circuit model required for hardware compilation. The \textbf{Genesis} \cite{chen2025genesis} compiler and that of Ref. \cite{decker2025symbolic} address hybrid Hamiltonian simulation but operate at a higher level of abstraction, lacking the granularity for general-purpose gate-level programming. \textbf{Bosonic Qiskit} \cite{stavenger2022bosonic} enables hybrid simulation by mapping qumodes onto qubit registers using a truncated Fock-space encoding ($n$ qubits for $2^n$ levels). However, this tight coupling between the representation and the simulation prevents true backend independence and native qumode abstraction.

\textbf{PennyLane} \cite{bergholm2018pennylane} provides a uniquely flexible architecture for hybrid extension. Its "wire" system simply addresses quantum objects while being agnostic to the type of the underlying quantum object, allowing it to handle qubits and qumodes. Furthermore, its symbolic gate definitions facilitate compilation and backend independence. Despite these strengths, standard PennyLane has limitations in a hybrid context: its measurement outcomes are typically represented as finite eigenvalue lists, which are incompatible with unbounded CV operators. Additionally, while PennyLane contains both DV and CV gates and routines, it lacks native hybrid gates, and its default simulators cannot perform joint hybrid state evolution.

No existing framework provides all the essential features for CV-DV quantum programming. Hybridlane aims to fill this gap by extending the PennyLane core library, resulting in an ergonomic frontend. Furthermore, by integrating Bosonic Qiskit for Fock-space simulation and the QSCOUT ion trap for hardware compilation, Hybridlane provides a unified frontend SDK for hybrid CV-DV quantum computing that satisfies all four architectural requirements identified in Table~\ref{tab:ecosystem-comparison}. A detailed comparison of trade-offs relative to each framework is given in Section~\ref{sec:related-work}.

\section{Design}
\label{sec:design}

Having established the need for a unified hybrid CV-DV frontend, we now present Hybridlane's core architecture and detail how it extends PennyLane to accomplish the above goals. Hybridlane's design is guided by three principles: (1) gate semantics are defined symbolically and decoupled from any particular matrix representation or simulation backend, enabling scalable circuit description; (2) wire types are inferred automatically and validated at compile time, catching errors before execution; and (3) the library extends PennyLane rather than replacing it, so that users retain access to the existing ecosystem of transforms, algorithms, and backends. We first discuss the wire type system (how Hybridlane distinguishes qubits from qumodes) and then proceed with the hybrid gate library and decompositions. Finally, we detail how Hybridlane handles the complexity of CV measurements and observables.

\subsection{Wire Type System and Inference}
\label{sec:wire_types}

PennyLane represents wires with hashable labels but carries no type information by default. While this flexibility enables CV-DV hybrid computation unlike many other libraries, it provides no static guarantees about operation compatibility. Hybridlane extends this wire system with a simple nominal type system comprising disjoint base types: $\{\mathtt{qubit}, \mathtt{qudit}(d), \mathtt{qumode}\}$ where $d=3, 4,\ldots$ is a finite integer dimension representing a qudit's Hilbert space dimensionality.

Wires are statically typed: once a wire's type is determined, it remains fixed throughout the circuit. This reflects the physical reality that quantum hardware operates on specific, incompatible modalities. Our type system enforces this incompatibility statically, preventing illegal operations such as applying qubit gates to qumode wires.

Requiring manual type annotations for every wire would be cumbersome for users. Instead, Hybridlane realizes automatic type inference using constraint propagation. Prior to device execution or circuit transformations, the inference algorithm performs a single forward pass through the circuit's instructions, determining wire types monotonically from operation signatures.

The type inference in Hybridlane exploits domain-specific properties of hybrid quantum circuits to be considerably simpler than type inference in general programming languages \cite{milner1978theory,carnier2024type}: (1)~wire types are disjoint with no subtyping, (2)~operations have monomorphic type signatures, and (3)~gate decompositions preserve type constraints.

\paragraph{Inference Procedure.}
The algorithm maintains a type environment, 
$$\Gamma: w \to \{\bot, \mathtt{qubit}, \mathtt{qudit}(d), \mathtt{qumode}\}$$ 
mapping wire labels to types, where $\bot$ represents an undetermined type. For each instruction in sequential order:

\begin{enumerate}
    \item \textbf{From gate operations:} The type signature is determined by known decompositions. First, we check if the operation has a known signature (e.g., Pauli gates require qubits). If not, we recursively analyze the gate's decomposition until reaching elementary gates with known signatures:
    \begin{itemize}
        \item If the gate inherits from \texttt{Operator} and defines a Pauli representation, all wires are constrained to $\mathtt{qubit}$.
        \item If the gate inherits from \texttt{CVOperator}, all wires are constrained to $\mathtt{qumode}$.
        \item For hybrid gates, a \texttt{Hybrid} mixin explicitly specifies each wire's type.
    \end{itemize}
    
    \item \textbf{From measurements:} Observable types constrain wire types. The algorithm decomposes symbolic observables recursively following the same procedure as gates. Additionally, qumodes can be identified from basis-specific measurements: \texttt{Continuous} basis sampling (position eigenstates) uniquely identifies qumodes, while \texttt{Discrete} basis (Fock states) is degenerate with qubit measurements and provides no constraint.
\end{enumerate}

Algorithm~1 summarizes the core inference loop. For each wire $w$ and required type $\tau$, the first determination wins ($\Gamma(w) := \tau$ when $\Gamma(w) = \bot$); a conflict raises a type error with diagnostic information. The \textsc{Signature} subroutine resolves an instruction's type constraints by checking, in order: (1)~known signatures (e.g., Pauli gates $\to$ qubit), (2)~\texttt{Hybrid} mixin annotations, or (3)~recursive decomposition into elementary gates with known signatures. Measurements follow the same resolution procedure applied to their symbolic observables. Since quantum gate decompositions are not polymorphic over wire types, no backtracking or fixpoint iteration is required.

\begin{figure*}[t]
\hrule\vspace{0.5em}
\textbf{Algorithm 1:} Wire Type Inference\label{alg:type-inference}
\vspace{0.3em}\hrule\vspace{0.5em}
\begin{algorithmic}[1]
\Require Quantum circuit tape $\mathcal{T} = (I_1, \ldots, I_n)$
\Ensure Type environment $\Gamma: w \to \{\mathtt{qubit}, \mathtt{qudit}(d), \mathtt{qumode}\}$
\State Initialize $\Gamma(w) \gets \bot$ for all wires $w$
\For{each instruction $I_k \in \mathcal{T}$}
    \For{each $(w, \tau)$ in $\Call{Signature}{I_k}$}
        \If{$\Gamma(w) = \bot$}
            \State $\Gamma(w) \gets \tau$
        \ElsIf{$\Gamma(w) \neq \tau$}
            \State \textbf{raise} TypeError($w$, $\Gamma(w)$, $\tau$)
        \EndIf
    \EndFor
\EndFor
\State \Return $\Gamma$
\end{algorithmic}
\vspace{0.3em}\hrule
\end{figure*}

\subsection{Hybrid gates and decompositions}
\label{sec:gates}

Following PennyLane's design, gates are defined symbolically by their parameters, wire arguments, and semantics (e.g. taking an adjoint or power, gate compositions). Therefore, gates are not coupled to a backend or a particular basis (like a matrix in Fock space), making Hybridlane's IR platform independent. A library of CV (Table \ref{tab:cv-gates}) and hybrid gates (Table \ref{tab:hybrid-gates}) is defined to match Ref. \cite{liu2026hybrid}, with the selected gates spanning the Gaussian, phase-space, Fock, and sideband ISAs.

There are two things to note about the gate definitions. Firstly, to remain compatible with PennyLane, we adopt the quantum information convention that $\ket{0} = \ket{g}$ is the $+1$ eigenstate of $Z$ and $\ket{1} = \ket{e}$ is the $-1$ eigenstate (cf. Ref. \cite{liu2026hybrid}, p. 17). Secondly, the $SWAP$ operation between bosonic modes is assigned its own operation \texttt{ModeSwap} instead of reusing the existing \texttt{Swap} gate to ensure that the wire arguments are not polymorphic.

\begingroup
\renewcommand{\arraystretch}{1.5}
\begin{table*}[htbp]
\centering
\small
\begin{tabular}{lll}
\toprule
\textbf{Name} & \textbf{Expression} & \textbf{Wires} \\
\midrule
Displacement & $D(\alpha) = \exp[\alpha\ad - \alpha^*a]$ & 1m \\
Rotation & $R(\theta) = \exp[-i\theta \hat{n}]$ & 1m \\
Fourier & $F = \exp[-i\frac{\pi}{2} \hat{n}]$ & 1m \\
Squeezing & $S(\zeta) = \exp[\frac{1}{2}(\zeta^* a^2 - \zeta (\ad)^2)]$ & 1m\\
Kerr & $K(\kappa) = \exp[-i \kappa \hat{n}^2]$ & 1m\\
Cubic Phase & $C(r) = \exp[-i r \hat{x}^3]$ & 1m\\
SNAP & $SNAP(\vec{\varphi}) = \sum_n e^{-i \varphi_n \ket{n}\bra{n}}$ & 1m\\
Beamsplitter & $BS(\theta, \varphi) = \exp[-i \frac{\theta}{2}(e^{i\varphi}\ad b + \text{h.c.})]$ & 2m\\
ModeSwap & $SWAP\ket{\psi_a, \psi_b} = \ket{\psi_b, \psi_a}$ & 2m \\
Two-mode Squeezing & $TMS(\xi) = \exp[\xi\ad b^\dagger - \text{h.c.}]$ & 2m\\
Two-mode Sum & $SUM(\lambda) = \exp[\frac{\lambda}{2}(a + \ad)(b^\dagger - b)]$ & 2m\\
\bottomrule
\end{tabular}
\caption{Qumode-only (CV) gates. The wires column gives the number of qumodes (m) each gate acts on.}
\label{tab:cv-gates}
\end{table*}
\endgroup

\begingroup
\renewcommand{\arraystretch}{1.5}
\begin{table*}[htbp]
\centering
\small
\begin{tabular}{llll}
\toprule
\textbf{Name} & \textbf{Expression} & \textbf{Wires} \\
\midrule
Conditional Rotation & $CR(\theta) = \exp[-i \frac{\theta}{2}Z\hat{n}]$ & 1q + 1m \\
Conditional Parity & $CP = \exp[-i\frac{\pi}{2}Z\hat{n}]$ & 1q + 1m\\
Conditional Displacement & $CD(\alpha) = \exp[Z(\alpha \ad - \alpha^* a)]$ & 1q + 1m\\
Conditional Squeezing & $CS(\zeta) = \exp[\frac{1}{2}Z (\zeta^* a^2 - \zeta (\ad)^2)]$ & 1q + 1m\\
SQR & $SQR(\vec{\theta}, \vec{\varphi}) = \sum_n R_{\varphi_n}(\theta_n) \otimes \ket{n}\bra{n}$ & 1q + 1m\\
Jaynes-Cummings (Red)\textsuperscript{$\dagger$} & $JC(\theta, \varphi) = \exp[-i\theta(e^{i\varphi}\sigma_- \ad + \text{h.c.})]$ & 1q + 1m\\
Anti-Jaynes-Cummings (Blue)\textsuperscript{$\dagger$} & $AJC(\theta, \varphi) = \exp[-i\theta(e^{i\varphi}\sigma_+ \ad + \text{h.c.})]$ & 1q + 1m\\
Rabi & $RB(\theta) = \exp[-iX (\theta \ad + \theta^*a)]$ & 1q + 1m\\
Conditional Beamsplitter & $CBS(\theta, \varphi) = \exp[-i\frac{\theta}{2}Z (e^{i\varphi}\ad b + \text{h.c.})]$ & 1q + 2m\\
Conditional Two-mode Squeezing & $CTMS(\xi) = \exp[Z (\xi \ad b^\dagger - \text{h.c.})]$ & 1q + 2m\\
Conditional Two-mode Sum & $CSUM(\lambda) = \exp[\frac{\lambda}{2}Z(a + \ad)(b^\dagger - b)]$ & 1q + 2m\\
\bottomrule
\end{tabular}
\caption{Hybrid CV-DV gates. The wires column gives the number of qubits (q) and qumodes (m) each gate acts on. \textsuperscript{$\dagger$}Red (blue) refers to the red (blue) sideband transition in AMO physics.}
\label{tab:hybrid-gates}
\end{table*}
\endgroup

In Table \ref{tab:decomps}, we give the known gate decompositions currently incorporated into Hybridlane. By leveraging PennyLane's graph decomposition system, Hybridlane can compile hybrid circuits to different platforms. In addition to these CV-DV identities, adjoints and powers of gates and symbolic operators are implemented, but we omit these for brevity. One innovation over PennyLane's existing system is the addition of the qubit-conditioned $C_q^Z(U)$ symbolic operator, which promotes any gate $U = e^{-i\theta G}$ to a qubit-conditioned version: $U \mapsto e^{-i\theta Z_q \otimes G}$. This operator enables automatic synthesis of qubit-controlled hybrid gates and gates conditioned on multiple qubits.

\begingroup
\renewcommand{\arraystretch}{1.5}
\begin{table*}[htbp]
\centering
\small
\begin{tabular}{lll}
\toprule
\# & \textbf{Decomposition rule} & \textbf{Reference} \\
\midrule
1 & $F = R(\pi/2)$ & \cite{liu2026hybrid}\\
2 & $SWAP_{i,j} = R_i(-\pi/2)R_j(-\pi/2)BS_{i.j}(\pi, 0)$ & \cite{liu2026hybrid}, \cite{crane2024hybrid}\\
3 & $CD_{q,m}(\alpha) = C^Z_q(D_m(\alpha))$ & \cite{liu2026hybrid}\\
4 & $CR_{q,m}(\theta) = C^Z_q(R_m(\theta/2))$ & \cite{liu2026hybrid}\\
5 & $CP_{q,m} = CR_{q,m}(\pi) = C^Z_q(F_m)$ & \cite{liu2026hybrid}\\
6 & $CS_{q,m}(\zeta) = C^Z_q(S_m(\zeta))$ & \cite{liu2026hybrid}\\
7 & $CBS_{q,m1,m2}(\theta, \varphi) = C^Z_q(BS_{m1,m2}(\theta, \varphi))$ & \cite{liu2026hybrid}\\
8 & $CTMS_{q,m1,m2}(\xi) = C^Z_q(TMS_{m1,m2}(\xi))$ & \cite{liu2026hybrid}\\
9 & $CSUM_{q,m1,m2}(\lambda) = C^Z_q(SUM_{m1,m2}(\lambda))$ & \cite{liu2026hybrid}\\
10 & $SNAP_m(\vec{\varphi}) = SQR_{anc, m}(-\vec{\pi}, \vec{\varphi})SQR_{anc,m}(\vec{\pi}, \vec{0})$ & \cite{liu2026hybrid}\\
11 & $
    C^Z_q(\exp[\hat{A}a^\dagger_m - \hat{A}^\dagger a_m]) = CP_{q,m} \exp[\hat{A}a^\dagger_m + \hat{A}^\dagger a_m] CP_{q,m}^\dagger$ & \cite{crane2024hybrid}\\
12 & $(C^Z_{q_1} \circ \dots \circ C^Z_{q_n})(U) = (\prod_{i=1}^{n-1} CNOT_{q_i,q_{i+1}}) C^Z_{q_n}(U) (\prod^1_{i=n-1} CNOT_{q_i, q_{i+1}})$ & \\
13 & $\ket{0}\bra{0}I + \ket{1}\bra{1}U = \sqrt{U}\sqrt{C^Z(U)}^\dagger$ & \\
\bottomrule
\end{tabular}
\caption{Currently implemented decompositions/equivalences for the gates}
\label{tab:decomps}
\end{table*}
\endgroup

In the table, Rule 11 is a generic decomposition identity that applies if the placeholder operator $\hat{A}$ satisfies $[\hat{A}, a^\dagger_m] = [\hat{A}, Z_q] = 0$, such as for $CD, CBS,$ and $CTMS$. Rule 12 synthesizes a more general unitary conditioned on the parity of $n$ qubits from a single qubit-conditioned gate and $2(n-1)$ CNOT gates. Rule 13 is another generic identity that enables synthesis of the usual qubit-controlled version of $U$ from the qubit-conditioned gate $C^Z(U)$. We remark that \cite{crane2024hybrid} has many more useful gate decompositions not yet implemented in Hybridlane because decompositions involving temporary ancillae qumodes require dynamic qumode allocation, which is not yet implemented. We leave this to future work.

\subsection{Measurements}
\label{sec:measurements}

PennyLane provides a robust measurement framework, supporting symbolic observables and diverse protocols such as expectation values, variances, basis-state histograms, and classical shadows. However, hybrid CV-DV computing introduces complexities, specifically unbounded observables and dual computational bases, that current PennyLane abstractions do not fully accommodate.

In standard DV systems, PennyLane represents measurement outcomes as a finite list of eigenvalues ($eigvals[i]$), mapping each computational basis state $i$ to an outcome. This approach fails for unbounded CV observables, such as position or momentum in bosonic Hamiltonians. To resolve this, Hybridlane introduces a \texttt{Spectral} mixin. Instead of a static list, this mixin allows an observable to define its eigenspectrum as a functional mapping, $f: \mathcal{B} \to \mathbb{R}$, from the basis space $\mathcal{B}$ to the corresponding eigenvalues. This functional approach ensures memory efficiency, as the system only processes a finite number of measurement shots regardless of the underlying spectrum's size. For hybrid systems, $\mathcal{B}$ can represent a composite space; for instance, an observable $Z\hat{x}$ operates on the space $\mathcal{B} = \mathbb{F}_2 \times \mathbb{R}$.

CV measurements typically occur in either the position basis (via homodyne readout) or the Fock basis (via photon-number counting). Capturing this nuance is vital for hardware compatibility and IR generation, as homodyne results require a \texttt{float} type while photon counting yields an \texttt{integer}. Hybridlane utilizes a \texttt{BasisSchema} object to track the measurement basis of each wire, categorized as $\{\mathtt{position}, \mathtt{discrete}\}$. Qubits and qudits are restricted to the $\mathtt{discrete}$ basis (yielding bits), while qumodes can be measured in the $\mathtt{position}$ basis (real-valued) or the $\mathtt{discrete}$ basis (integer-valued).

The framework employs an inference algorithm to determine the appropriate basis based on the observables being measured. While the \texttt{Spectral} mixin currently guides this inference by specifying the most efficient basis for diagonalization, future iterations will refine this logic for more complex symbolic observables. For cases where the user would like to specify a low-level measurement, there exists a function to sample a particular computational basis.

While Hybridlane expands the scope of hybrid measurement, challenges remain in achieving full parity with PennyLane’s DV suite. Particularly, mid-circuit measurement (MCM) for qumodes is not yet supported. PennyLane allows conditional branching on the \textit{value} of a MCM, but to ensure a finite number of conditional execution branches, future CV-MCM implementations could utilize boolean predicates (e.g., branching only if $x > \text{threshold}$).

\section{Backend Integration}
\label{sec:backends}

Hybridlane inherits PennyLane's device architecture, which cleanly separates high-level circuit description from low-level execution. By implementing PennyLane's \texttt{Device} interface, Hybridlane backends provide a unified abstraction that enables the same circuit to target different execution environments. The architecture enables future integrations with additional backends such as photonic hardware, superconducting cavity systems, or alternative simulators.

Devices in Hybridlane largely follow those in PennyLane and therefore will be familiar to existing PennyLane developers. The main difference is that in Hybridlane, the type inference algorithm can be run as part of the device's \texttt{QuantumTape} preprocessing pipeline to both validate circuit structure and inform circuit transforms.

Currently, two devices are available:

\begin{itemize}
    \item \texttt{bosonicqiskit.hybrid}: A simulator device that translates Hybridlane circuits to Bosonic Qiskit for simulation in Fock space using qubits.

    \item \texttt{sandiaqscout.hybrid}: A compilation-only device that serves as a target for the Sandia QSCOUT ion trap. Compiled circuits are exported to JAQAL for execution on hardware.
\end{itemize}

\subsection{Bosonic Qiskit}
\label{subsec:bosonic-qiskit}

The \texttt{bosonicqiskit.hybrid} device bridges Hybridlane with the Bosonic Qiskit simulator \cite{stavenger2022bosonic}, enabling the classical simulation of hybrid circuits through symbolic decomposition. To simulate hybrid CV-DV systems, Bosonic Qiskit truncates each qumode to a finite Fock-state subspace represented by a set of qubits. Every gate is subsequently defined by its sparse matrix representation in the Fock basis, allowing the resulting qubit circuit to be executed via the Qiskit Aer backend.

To utilize this device within the Hybridlane framework, the user must first define the Fock truncation levels for each qumode. Because Bosonic Qiskit maps these states to qubits, truncation levels are restricted to powers of two. The integration process follows a structured pipeline:

\begin{enumerate}
    \item \textbf{Unrolling}: The Hybridlane program is flattened into a sequential list of quantum gates, internally as the PennyLane \texttt{QuantumScript} IR.

    \item \textbf{Decomposition}: Gates are decomposed into the native Bosonic Qiskit gate set.

    \item \textbf{Translation}: The \texttt{QuantumScript} is inspected by the Hybridlane inference algorithms, and the typed quantum circuit is mapped to a Bosonic Qiskit \texttt{CVCircuit} object.

    \item \textbf{Diagonalization}: Any gates required to rotate measured observables into their respective eigenbases (specifically the $Z$ and Fock bases) undergo the same decomposition and translation steps.
\end{enumerate}

The device supports both analytic and finite-shot measurement protocols. For analytic measurements, Hybridlane constructs the Fock representation of an observable directly from its symbolic expression tree. This representation is then applied to the statevector obtained from the simulation to derive the result. In finite-shot mode (\texttt{shots != None}), observables are first diagonalized into their computational bases; Hybridlane then collects the resulting basis-state samples and performs the necessary post-processing to return the final measurement values.

The following example illustrates the typical usage pattern:

\begin{lstlisting}[]
dev = qml.device("bosonicqiskit.hybrid", max_fock_level=8)

@qml.qnode(dev)
def circuit(alpha):
    qml.X(0)
    hqml.D(alpha, 0, "m")
    hqml.JC(0.5, 0, [0, "m"])
    return hqml.expval(hqml.N("m"))
    
result = circuit(0.3)
\end{lstlisting}

\subsection{Sandia QSCOUT Ion Trap}
\label{subsec:sandia}

The \texttt{sandiaqscout.hybrid} device enables the compilation of Hybridlane circuits into the JAQAL IR~\cite{morrison2020jaqal} for execution on the QSCOUT ion-trap platform at Sandia National Laboratories~\cite{clark2021qscout}.

QSCOUT is a trapped-ion quantum processor where $n$ ionic qubits naturally couple to $2n$ collective motional modes (qumodes). These motional modes are organized by spatial "manifold" (labeled 0 and 1) and mode index. In this architecture, mode 0 represents the center-of-mass (COM) mode, where all ions oscillate in phase, while modes 1 through $n-1$ represent tilt modes characterized by relative phase oscillations. Hybridlane adopts a specific wire-naming convention to map hardware qumodes: \texttt{"m\{manifold\}i\{index\}"}. For example, \texttt{"m1i1"} identifies the first tilt mode on manifold 1, while qubits are identified by standard integer indices $0, \dots, n-1$.

As the QSCOUT platform is under active development, the Hybridlane device incorporates some experimental hardware constraints:
\begin{itemize}
    \item \textbf{Qumode availability}: By default, COM modes are disabled due to their high heating rates, though they may be manually enabled by the user. Moreover, the number of qubits in the circuit constrains the number of available qumodes.
    
    \item \textbf{Fixed-qumode couplings}: Certain hybrid gates are currently restricted to specific qumodes. For instance, the native beamsplitter gate couples the lowest-order tilt modes between manifolds 0 and 1.
\end{itemize}

To streamline programming, the device validates operations prior to compilation to ensure adherence to hardware constraints. It also features an optional virtual wire allocation mode, which dynamically maps virtual qumode wires (not labeled according to the hardware convention above) to physical hardware modes to satisfy gate-specific connectivity requirements.

The compilation workflow largely parallels that of the \texttt{bosonicqiskit.hybrid} device, with the addition of the virtual wire allocation step following gate translation. However, as the QSCOUT platform does not currently support direct cloud-based submission, the device serves as a compilation target rather than a real-time execution backend. Circuits are instead exported to the JAQAL IR, allowing for manual deployment on the hardware. In Section~\ref{sec:applications} we give an example workflow of defining a circuit, validating it on a classical simulator, and then exporting it to JAQAL using the \texttt{sandiaqscout.hybrid} device.

The Hybridlane device is intended to co-evolve with the QSCOUT hardware; consequently, the current gate equivalences and constraints are subject to change. Future work will focus on integrating pulse-level control, a capability natively supported by the JAQAL language~\cite{lobser2023}.

\subsection{OpenQASM Intermediate Representation}
\label{subsec:openqasm}

\paragraph{Language extensions}
Hybridlane extends OpenQASM 3.0~\cite{cross2022openqasm} to support continuous-variable and hybrid quantum circuits, providing an intermediate representation for circuit interchange and archival. Our format differs from the existing CVDV-QASM \cite{chen2025genesis} because (1)~we introduce new keywords to preserve important typing information, and (2)~our format does not contain the Pauli string instructions tailored to Hamiltonian simulation.

Three keywords are introduced to support hybrid circuits:
\begin{enumerate}
\item \textbf{\texttt{qumode}}: Declares qumode registers analogous to \texttt{qubit}.

\item \textbf{\texttt{measure\_x}}: Homodyne (position) measurement, returning floating-point values.

\item \textbf{\texttt{measure\_n}}: Fock (photon number) measurement, returning unsigned integers.
\end{enumerate}

These extensions enable explicit type information and measurement basis specification in the intermediate representation. Furthermore, in quantum \texttt{gate} definitions and \texttt{defcal} blocks defined for virtual quantum arguments, we type each argument with \texttt{qubit} or \texttt{qumode}.

Hybridlane provides a standard gate library defining continuous-variable and hybrid gates using OpenQASM's \texttt{gate} and \texttt{defcal} syntax. The gate library is included with \texttt{include "cvstdgates.inc";} statements in exported circuits. For example, this is the definition for the $CD$ gate, using the OpenQASM modifiers to express it as $CD(\alpha) = \ket{0}\bra{0}D(\alpha) + \ket{1}\bra{1}D^\dagger(\alpha)$.

\begin{lstlisting}[language=openqasm]
// Conditioned-displacement gate
gate cv_cd(r, phi) qubit q, qumode m {
    negctrl @ cv_d(r, phi) q, m;
    ctrl @ inv @ cv_d(r, phi) q, m;
}
\end{lstlisting}

\paragraph{Usage example}
Hybridlane provides a \texttt{to\_openqasm()} function to export circuits to our OpenQASM language. Consider the following snippet which prepares the even cat state $\ket{\mathcal{C}_\alpha} \approx \ket{\alpha} + \ket{-\alpha}$ using the non-Abelian QSP protocol \cite{singh2025towards} and reads out the photon number of the qumode:

\begin{lstlisting}
dev = qml.device("bosonicqiskit.hybrid", max_fock_level=8)

@qml.qnode(dev)
def circuit(alpha):
    hqml.SqueezedCatState(alpha, np.pi / 2, wires=["q", "m"])
    return hqml.expval(hqml.N("m"))

hqml.to_openqasm(circuit, precision=4)(4)
\end{lstlisting}

This results in the output (for $\alpha = 4$)

\begin{lstlisting}[language=openqasm,basicstyle=\scriptsize\ttfamily]
OPENQASM 3.0;
include "stdgates.inc";
include "cvstdgates.inc";

qubit[1] q;
qumode[1] m;

def state_prep() {
    reset q;
    reset m;
    h q[0];
    cv_cd(4.0000, 0.0000) q[0], m[0];
    h q[0];
    sdg q[0];
    h q[0];
    cv_cd(0.0982, 1.5708) q[0], m[0];
    h q[0];
    sdg q[0];
    h q[0];
    cv_cd(0.0982, 3.1416) q[0], m[0];
    h q[0];
    s q[0];
    s q[0];
}

state_prep();
uint c0 = measure_n m[0];
\end{lstlisting}

Here one can see the CV-DV language extensions. Both the qubit and hybrid standard gate libraries are imported, and the quantum registers \texttt{q} and \texttt{m} are typed. Finally, the photon number readout of the qumode is stored in the classical variable \texttt{c0} at machine precision.

\section{Applications and Demonstrations}
\label{sec:applications}

As we envision Hybridlane to unify the software ecosystem of CV-DV quantum computing, we demonstrate its utility through two examples. The first is simulation of quantum phase estimation (QPE) of a hybrid CV-DV system, leveraging the decompositions built in to the library and existing PennyLane templates. The second one demonstrates reusing the same circuit across multiple devices by calibrating an ion trap.

\subsection{Quantum Phase Estimation on Dispersive Hamiltonian}

We apply QPE to the dispersive Hamiltonian
\begin{equation}
    H = \omega_r \hat{n} - \frac{\omega_q}{2}Z - \frac{\chi}{2}Z \hat{n}.
    \label{eq:dispersive_ham}
\end{equation}
Because the terms are mutually commuting, the time evolution operator is $U(t) = R_Z(-\omega_q t)R(\omega_r t) CR(-\chi t)$. This can be expressed as an operation in Hybridlane as follows:
\begin{lstlisting}
class Evo(Operation, Hybrid):
    num_wires = 2
    num_params = 1

    type_signature = (hqml.sa.Qubit(), hqml.sa.Qumode())
    resource_keys = set()

    def __init__(self, t: float, omega_r=1, omega_q=-1, chi=0.1, wires=None, id=None):
        self.hyperparameters.update(
            {"omega_r": omega_r, "omega_q": omega_q, "chi": chi}
        )
        super().__init__(t, wires=wires, id=id)
\end{lstlisting}
This largely follows PennyLane syntax, so we focus on the new aspects of Hybridlane. Particularly, as this is a CV-DV Hamiltonian, the gate additionally inherits from \texttt{Hybrid} to help inform the type inference algorithm. The \texttt{Hybrid} mixin adds the property \texttt{type\_signature}, which here tells Hybridlane that the first wire (\texttt{wires[0]}) is a qubit and the second wire (\texttt{wires[1]}) is a qumode. If this was omitted, Hybridlane would traverse the decomposition and determine the wire types from its knowledge of the $R_Z$, $R$ and $CR$ gates.

\begin{figure*}[htbp]
    \centering
    \includegraphics[width=\linewidth]{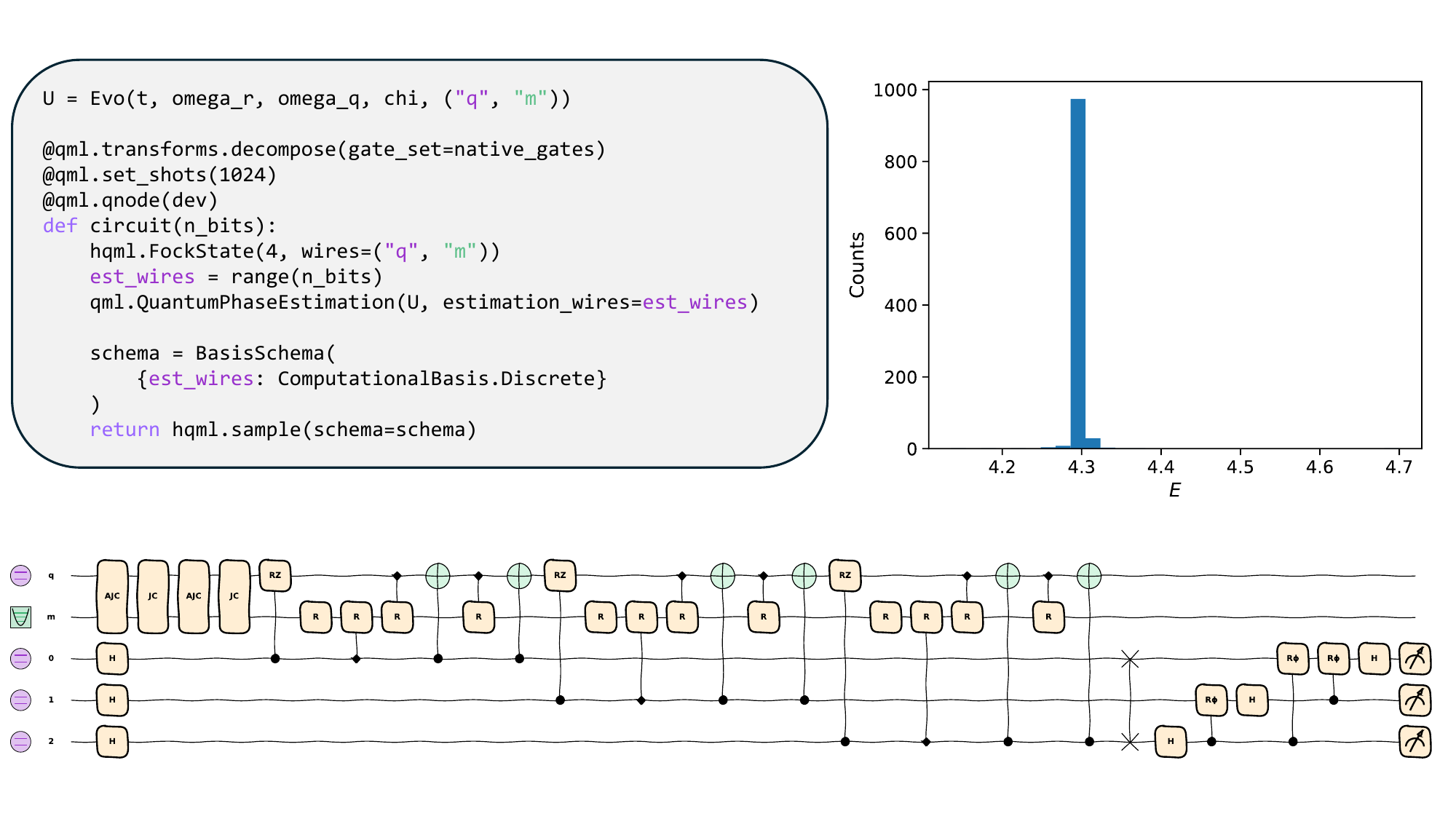}
    \caption{Overview of performing quantum phase estimation on a simple hybrid Hamiltonian in (\ref{eq:dispersive_ham}). \textbf{(Left)} Code block expressing the circuit. After using a Hybridlane template to prepare the test eigenstate $\ket{0,4}$, the PennyLane QPE template is applied, and finally the qubits are read out in their discrete computational basis ($Z$). The wire names are colored by type. \textbf{(Right)} Histogram of the eigenvalue distribution with 10 estimation qubits and 1024 shots. \textbf{(Bottom)} The decomposition of the circuit into the native gate set of the Bosonic Qiskit simulator.}
    \label{fig:qpe-overall}
\end{figure*}

The decomposition in terms of primitive gates uses PennyLane's new graph decomposition system:
\begin{lstlisting}
@qml.register_resources({qml.RZ: 1, hqml.R: 1, hqml.CR: 1})
def _evo_decomp(t, wires, omega_r, omega_q, chi):
    qml.RZ(-omega_q * t, wires[0])
    hqml.R(omega_r * t, wires[1])
    hqml.CR(-chi * t, wires)

qml.add_decomps(Evo, _evo_decomp)
qml.add_decomps("Pow(Evo)", pow_rotation)
\end{lstlisting}
The final line just expresses that $U^k(t) = U(kt)$, which is the \texttt{pow\_rotation} symbolic decomposition rule. With this, we can express the QPE circuit:
\begin{lstlisting}
U = Evo(t, omega_r, omega_q, chi,
    wires=("q", "m"))

dev = qml.device("bosonicqiskit.hybrid",
    max_fock_level=8)

@qml.transforms.decompose(gate_set=native_gates)
@qml.set_shots(1024)
@qml.qnode(dev)
def circuit(n_bits: int):
    hqml.FockState(4, wires=("q", "m"))
    est_wires = range(n_bits)
    qml.QuantumPhaseEstimation(U,
        estimation_wires=est_wires)

    schema = BasisSchema(
        {est_wires: ComputationalBasis.Discrete}
    )
    return hqml.sample(schema=schema)
\end{lstlisting}
Here we prepare the definite eigenstate $\ket{0, 4}$ and measure its eigenvalue. Behind the scenes, several decomposition identities (Section~\ref{sec:gates}) are applied to synthesize the controlled unitary $c(U^k)$ required for QPE. For 3 estimation qubits, the decomposition in the Bosonic Qiskit gate set is shown in Fig.~\ref{fig:qpe-overall}. Using 10 estimation qubits and postprocessing the bitstrings, we obtain a distribution tightly concentrated on $E = 4.3013$ with parameters $t=1, \omega_r = 1, \omega_q = -1, \chi=0.1$.

\subsection{Ion Trap Calibration Workflow}

\begin{figure*}[htbp]
    \centering
    \includegraphics[width=\linewidth]{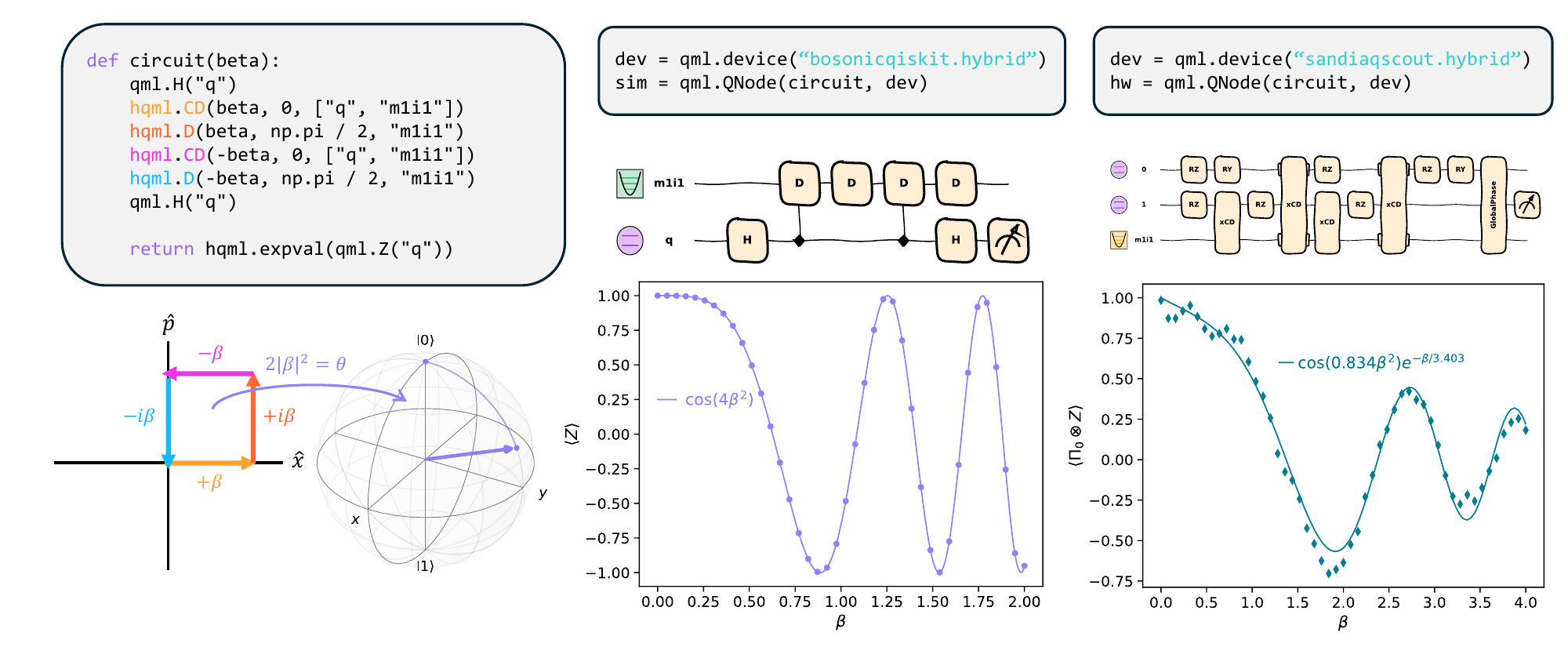}
    \caption{Overview of performing $CD$ gate calibration using Hybridlane. \textbf{(Left)} The calibration circuit maps the area of a loop in oscillator phase-space into a rotation on an ancilla qubit \cite{liu2026hybrid}. Like in PennyLane, we specify the circuit instructions as a Python function to be reused across devices. \textbf{(Center)} Simulating the circuit using Bosonic Qiskit allows verification of its behavior prior to running on hardware. \textbf{(Right)} By switching Hybridlane devices, the same circuit is transformed to run on the Sandia QSCOUT ion trap. Because part of the decomposition process uses an auxiliary qubit (\texttt{0}) to realize the $D$ gates, the output is postselected on the auxiliary qubit ending in state $\ket{0}$.}
    \label{fig:cd-overview}
\end{figure*}

An approach to calibrating the native spin-dependent force (SDF) gate on QSCOUT uses Ramsey interferometry: leveraging phase-kickback, a conditional displacement creates qubit-phase sensitivity to motional mode displacement \cite{liu2026hybrid}. Hybridlane's multi-backend workflow (Fig.~\ref{fig:cd-overview}) validates the circuit with Bosonic Qiskit simulation (producing expected oscillations), then compiles to JAQAL for hardware execution. The QSCOUT device automatically decomposes to the native gate set and validates gate support.

The core routine of the calibration circuit is
\begin{equation}
    U(\beta) = D(-i\beta)CD(-\beta)D(i\beta)CD(\beta).
\end{equation}
As seen in Fig.~\ref{fig:cd-overview}, this produces a counter-clockwise (clockwise) loop in phase space when applied to the qubit in state $\ket{0}$ ($\ket{1}$), imparting a phase of $e^{2i|\beta|^2}$ ($e^{-2i|\beta|^2}$). By applying the full circuit $H U(\beta)H$ to the initial state $\ket{0,0}$, an $R_X(\theta)$ gate on the qubit is realized where $\theta = 4|\beta|^2$. The expectation value is then $\langle Z \rangle = \cos(4|\beta|^2)$.

We begin by validating our circuit definition using the \texttt{bosonicqiskit.hybrid} device with the following code block:
\begin{lstlisting}
def circuit(beta):
    qml.H("q")
    hqml.CD(beta, 0, ["q", "m1i1"])
    hqml.D(beta, np.pi / 2, "m1i1")
    hqml.CD(-beta, 0, ["q", "m1i1"])
    hqml.D(-beta, np.pi / 2, "m1i1")
    qml.H("q")
    return hqml.expval(qml.Z("q"))

dev = qml.device("bosonicqiskit.hybrid",
    max_fock_level=16)
sim = qml.QNode(circuit, dev)
betas = np.linspace(0, 2, 40)
expvals = list(map(sim, betas))
\end{lstlisting}
As this is a benchmark, we specify a particular physical qumode for the ion trap using the notation in Section~\ref{subsec:sandia}. This has no effect on the Bosonic Qiskit simulator. The middle region of Fig.~\ref{fig:cd-overview} confirms that our circuit produces the expected oscillation.

Next we compile the circuit to the QSCOUT ion trap by binding the \texttt{circuit} function to a different \texttt{QNode}. The qumode displacements $D$ are realized through dynamic qubit allocation: $D \rightarrow CD\ket{0}$, and the decomposition adds single qubit gates to transform the $CD$ gates to the native $xCD$ gates. After decomposition, the circuit is exported to the native JAQAL language of the ion trap. Here we fix $\beta=0.5$ for brevity, as each value of $\beta$ would result in a new \texttt{subcircuit} block.
\begin{lstlisting}
dev = qml.device("sandiaqscout.hybrid",
    optimize=True, n_qubits=2)
hw = qml.set_shots(qml.QNode(circuit, dev), 1024)
ir = to_jaqal(hw, level="device",
    precision=5)(0.5)
\end{lstlisting}

The JAQAL code is
\begin{lstlisting}[language=jaqal]
from Calibration_PulseDefinitions.QubitBosonPulses usepulses *

register q[2]

subcircuit {
	Rz q[1] 6.2832
	xCD q[1] 1 1 0.5 0.0
	Rz q[1] 6.2832
	Rz q[0] 3.1416
	Ry q[0] 1.5708
	xCD q[0] 1 1 0.0 0.5
	Rz q[0] 6.2832
	xCD q[1] 1 1 -0.5 -0.0
	Rz q[1] 6.2832
	xCD q[0] 1 1 -0.0 -0.5
	Rz q[0] 3.1416
	Ry q[0] 1.5708
}
\end{lstlisting}

We show the output of the QSCOUT hardware on the right side of Fig.~\ref{fig:cd-overview} with $2000$ shots per $\beta$. Because the ancilla qubit \texttt{0} is used only to realize the (unconditioned) displacement on the qumode, the output was post-selected to remove shots where the ancilla qubit ended in state $\ket{1}$, which would signify an error. The median rejection rate across all values of $\beta$ was $2.6\%$. The sampled points were then fit to a decaying cosine function to allow for hardware noise, giving $\cos(0.834\beta^2)e^{-\beta/3.403}$. The difference of $\sqrt{4/0.834} \approx 2.19$ in the oscillation frequency has been used to improve the QSCOUT gate set, demonstrating utility of the unified Hybridlane workflow.

\section{Related Work}
\label{sec:related-work}

Section~\ref{sec:background} surveyed the quantum software ecosystem broadly; here we provide a focused comparison between Hybridlane and the most closely related frameworks, acknowledging trade-offs inherent in our design choices.

\paragraph{Comparison to PennyLane.}
PennyLane provides automatic differentiation for variational quantum algorithms, enabling gradient-based parameter optimization through parameter-shift rules and backpropagation. Hybridlane currently lacks automatic differentiation support, as Bosonic Qiskit does not expose differentiable interfaces. Additionally, PennyLane's Catalyst framework offers just-in-time compilation via MLIR for performance-critical workloads, which Hybridlane does not yet support. However, Hybridlane's native qumode types and comprehensive hybrid gate library go beyond PennyLane's DV-or-CV approach.

\paragraph{Comparison to Strawberry Fields.}
Strawberry Fields offers highly optimized backends for CV-only quantum computing, including fast Gaussian simulations and specialized photonic hardware compilation. Its Gaussian backend can efficiently simulate hundreds of qumodes when restricted to Gaussian operations (linear optics, displacements, squeezing). Hybridlane currently lacks a dedicated Gaussian simulator, relying on general Fock-space simulation via Bosonic Qiskit. For purely CV workflows without qubit coupling, Strawberry Fields may offer better performance. However, for hybrid applications requiring qubit-qumode interactions (ion traps, cavity QED), Hybridlane's architecture is essential.

\paragraph{Comparison to Bosonic Qiskit.}
Hybridlane depends on Bosonic Qiskit for classical simulation, rather than providing a custom simulator. This introduces an external dependency and limits simulation performance to Bosonic Qiskit's capabilities. However, this design choice reflects Hybridlane's intentional focus as a frontend library. By separating circuit description from execution, Hybridlane achieves scalability and backend flexibility and enables circuit transforms that Bosonic Qiskit alone cannot provide. The frontend-backend separation also enables future integration with alternative simulators (e.g., GPU-accelerated Fock simulators, tensor network simulators) without API changes.

\paragraph{Other frameworks.}
Piquasso is a photonic quantum simulation platform supporting both discrete Fock states and continuous-variable representations, with differentiable JAX and TensorFlow backends. While it offers gradient-based optimization that Hybridlane currently lacks, Piquasso is primarily a simulator rather than a backend-independent frontend SDK, and does not target hybrid CV-DV hardware platforms such as ion traps. QuTiP provides general-purpose quantum simulation at the Hamiltonian and density matrix level, supporting hybrid systems through tensor products, but does not adhere to the gate-based circuit model and therefore occupies a complementary niche. MrMustard offers differentiable phase-space and Fock-space representations for photonic circuits, but similarly focuses on simulation rather than circuit description and backend dispatch.

\section{Conclusion}
\label{sec:conclusion}

Hybrid continuous-discrete variable quantum computing represents a promising frontier that combines the robust digital encoding of qubits with the large Hilbert spaces and analog capabilities of qumodes. Hardware platforms spanning trapped ions, superconducting circuits, and photonic systems naturally realize hybrid architectures, yet the software ecosystem has lagged behind, forcing researchers to work with fragmented toolchains or simulation-coupled representations that limit scalability and portability.

By providing a unified frontend for hybrid quantum programming, Hybridlane lowers the barrier to entry for researchers exploring CV-DV algorithms. The separation of circuit description from execution enables rapid prototyping without simulation overhead, while static type safety prevents costly errors on hardware backends. The OpenQASM extensions establish a foundation for interoperability, enabling circuit sharing and archival across research groups. Built on PennyLane's plugin architecture, Hybridlane immediately integrates with an established ecosystem while extending its capabilities to hybrid systems, offering a familiar API for PennyLane users.

Several limitations scope the current work. Hybridlane does not yet support automatic differentiation, as its simulation backend (Bosonic Qiskit) does not expose differentiable interfaces; variational hybrid algorithms therefore require gradient-free or finite-difference optimizers. Simulation is limited to Fock-space truncation with power-of-two cutoffs, and no dedicated Gaussian or noisy simulator is provided. The hardware backend is restricted to a single platform (QSCOUT ion trap), and mid-circuit measurement for qumodes is not yet implemented. Finally, Hybridlane is in alpha, and its API may evolve as the hybrid CV-DV ecosystem matures.

We aim to address these limitations through community collaboration. Near-term priorities include differentiable simulation for variational algorithms, symbolic Hamiltonian simulation compiler integration, pulse-level control, and extended backend support for photonic and superconducting cavity platforms.

As hybrid CV-DV hardware continues to advance across trapped-ion, superconducting, and photonic platforms, the need for a unified software layer will only grow. Hybridlane provides a foundation for this layer: an open, extensible SDK where the same circuit can move from whiteboard to simulator to hardware with minimal friction. We invite the community to join us in building out this ecosystem and welcome contributions.

\section*{Acknowledgments}
This work was supported by the U.S. Department of Energy under Grant No. DE-FOA-0003265. Pacific Northwest National Laboratory is operated by Battelle Memorial Institute for the U.S. Department of Energy under Contract No. DE-AC05-76RL01830. We thank the Bosonic Qiskit developers for simulation backend support, Christopher Yale and the Sandia QSCOUT team for hardware access and collaboration, and Erik Lentz for feedback and discussion. This research was supported in part by the U.S. Department of Energy, Office of Science, Office of Advanced Scientific Computing Research Quantum Testbed Program.

\bibliographystyle{quantum}
\bibliography{references}

\end{document}